\documentclass[twocolumn,floatfix,superscriptaddress,showpacs]{revtex4}
\usepackage{graphicx}
\begin{document}
\title{Exponential sensitivity to dephasing of electrical conduction
through a quantum dot} 
\author{J. Tworzyd{\l}o}
\affiliation{Instituut-Lorentz, Universiteit Leiden, P.O. Box 9506,
2300 RA Leiden, The Netherlands} 
\affiliation{Institute of Theoretical
Physics, Warsaw University, Ho\.{z}a 69, 00--681 Warsaw, Poland}
\author{A. Tajic} 
\affiliation{Instituut-Lorentz, Universiteit Leiden,
P.O. Box 9506, 2300 RA Leiden, The Netherlands} 
\author{H. Schomerus}
\affiliation{Max-Planck-Institut f\"{u}r Physik komplexer Systeme,
N\"{o}thnitzer Str.\ 38, 01187 Dresden, Germany}
\author{P. W. Brouwer} 
\affiliation{Laboratory of Atomic and Solid
State Physics, Cornell University, Ithaca, NY 14853--2501, USA}
\author{C.W.J. Beenakker} 
\affiliation{Instituut-Lorentz, Universiteit
Leiden, P.O. Box 9506, 2300 RA Leiden, The Netherlands} \date{July
2004}
\begin{abstract}
According to random-matrix theory, interference effects in the
conductance of a ballistic chaotic quantum dot should vanish
$\propto(\tau_{\phi}/\tau_{D})^{p}$ when the dephasing time
$\tau_{\phi}$ becomes small compared to the mean dwell time
$\tau_{D}$. Aleiner and Larkin have predicted that the power law
crosses over to an exponential suppression
$\propto\exp(-\tau_{E}/\tau_{\phi})$ when $\tau_{\phi}$ drops below
the Ehrenfest time $\tau_{E}$. We report the first observation of this
crossover in a computer simulation of universal conductance
fluctuations. Their theory also predicts an exponential suppression
$\propto\exp(-\tau_{E}/\tau_{D})$ in the absence of dephasing ---
which is {\em not\/} observed. We show that the effective
random-matrix theory proposed previously for quantum dots without
dephasing explains both observations.
\end{abstract}
 \maketitle

An instructive way to classify quantum interference effects in
mesoscopic conductors is to ask whether they depend exponentially or
algebraically on the dephasing time $\tau_{\phi}$. The Aharonov-Bohm
effect is of the former class, while weak localization (WL) and
universal conductance fluctuations (UCF) are of the latter class
\cite{Alt91,Imr02}. It is easy enough to understand the difference: On
the one hand, Aharonov-Bohm oscillations in the magnetoconductance of
a ring require phase coherence for a certain minimal time $t_{\rm
min}$ (the time it takes to circulate once along the ring), which
becomes exponentially improbable if $\tau_{\phi}<t_{\rm min}$. On the
other hand, WL and UCF in a disordered quantum dot originate from
multiple scattering on a broad range of time scales, not limited from
below, and the superposition of exponents with a range of decay rates
amounts to a power law decay.

In a seminal paper \cite{Ale96}, Aleiner and Larkin have argued that
ballistic chaotic quantum dots are in a class of their own. In these
systems the Ehrenfest time $\tau_{E}$ introduces a lower limiting time
scale for the interference effects, which are exponentially suppressed
if $\tau_{\phi}<\tau_{E}$. The physical picture is that electron wave
packets in a chaotic system can be described by a single classical
trajectory for a time up to $\tau_{E}$ \cite{Zas81}. Both WL and UCF,
however, require that a wave packet splits into partial waves which
follow different trajectories before interfering. Only the fraction
$\exp(-\tau_{E}/\tau_{\phi})$ of electrons which have not yet dephased
at time $\tau_{E}$ can therefore contribute to WL and UCF.

The WL correction $\Delta G=\langle G\rangle-G_{\rm cl}$ is the
deviation of the ensemble averaged conductance $\langle G\rangle$ (in
zero magnetic field) from the classical value $G_{\rm cl}=N/2$. (We
measure conductances in units of $2e^{2}/h$ and assume an equal number
of modes $N\gg 1$ in the two leads that connect the quantum dot to
electron reservoirs.) The WL correction according to random-matrix
theory (RMT),
\begin{equation}
\Delta G_{\rm
RMT}=-{\textstyle\frac{1}{4}}(1+\tau_{D}/\tau_{\phi})^{-1},
\label{deltaGRMT}
\end{equation}
has a power law suppression $\propto\tau_{\phi}/\tau_{D}$ when
$\tau_{\phi}$ becomes smaller than the mean dwell time $\tau_{D}$ in
the quantum dot \cite{Bro97}. Similarly, RMT predicts for the UCF a
power law suppression $\propto(\tau_{\phi}/\tau_{D})^{2}$ of the
mean-squared sample-to-sample conductance fluctuations
\cite{Bro97,Bar95},
\begin{equation}
{\rm Var}\,G_{\rm RMT}=\frac{1}{8\beta}(1+\tau_{D}/\tau_{\phi})^{-2},
\label{VarGRMT}
\end{equation}
with $\beta=2$ $(1)$ in the presence (absence) of a
time-reversal-symmetry-breaking magnetic field.

Aleiner and Larkin have calculated the $\tau_{E}$-dependence of the WL
correction, with the result \cite{Ale96}
 \begin{equation}
\Delta G=e^{-\tau_{E}/\tau_{\phi}}e^{-2\tau_{E}/\tau_{D}}
\Delta G_{\rm RMT}.
\label{deltaGAL}
\end{equation}
The two exponential suppression factors in Eq.\ (\ref{deltaGAL})
result from the absence of interfering trajectories for times below
$\tau_{E}$. The first factor $\exp(-\tau_{E}/\tau_{\phi})$ accounts
for the loss by dephasing and the second factor
$\exp(-2\tau_{E}/\tau_{D})$ accounts for the loss by escape into one
of the two leads. The UCF are expected to be suppressed similarly.

The physical picture presented by Aleiner and Larkin is simple and
supported by two independent calculations \cite{Ale96,Ada03}. And yet,
it has been questioned as a result of some very recent computer
simulations of UCF \cite{Jac04,Two04a} and WL \cite{Two04b} in the
absence of dephasing. The expected exponential reduction of quantum
interference effects due to escape into the leads was not observed. In
fact, both WL and UCF were found to be completely independent of
$\tau_{E}$, even though the simulations extended to system sizes for
which $\tau_{E}$ was well above $\tau_{D}$. To explain these negative
results, Jacquod and Sukhorukov \cite{Jac04} invoked the {\em
effective\/} RMT of Silvestrov et al.\ \cite{Sil03a}. In that approach
the electrons with dwell times $>\tau_{E}$ are described by RMT with
an effective number $N_{\rm eff}=N\exp(-\tau_{E}/\tau_{D})$ of
modes. Then no $\tau_{E}$-dependence is expected as long as $N_{\rm
eff}\gg 1$ --- even if $\tau_{E}\gg\tau_{D}$.

Since the predicted exponential reduction factor due to escape into
the leads has not appeared in the simulations, it is natural to ask
about the factor $\propto\exp(-\tau_{E}/\tau_{\phi})$ due to
dephasing. Does it exist? An experimental study of two-dimensional
(2D) weak localization has concluded that it does \cite{Yev00}, but
since leads play no role in 2D these experiments can not really
resolve the issue. In the absence of experiments on the
zero-dimensional geometry of a quantum dot, we have used computer
simulations to provide an answer. We find that a relatively small
amount of dephasing is sufficient to introduce a marked
$\tau_{E}$-dependence of the UCF. Our observation can be explained by
incorporating dephasing into the effective RMT. We find that
\begin{eqnarray}
&&\Delta G=e^{-\tau_{E}/\tau_{\phi}}\Delta G_{\rm RMT},
\label{DeltaGresult}\\ &&{\rm Var}\,G=e^{-2\tau_{E}/\tau_{\phi}}\,{\rm
Var}\,G_{\rm RMT}, \label{VarGresult}
\end{eqnarray}
and show that Eq.\ (\ref{VarGresult}) provides a
fitting-parameter-free description of the numerical data.

We have introduced a dephasing lead \cite{But86} in the kicked
rotator, which is the same dynamical system studied in Refs.\
\cite{Jac04,Two04a,Two04b,Tia04} in the absence of dephasing. The
kicked rotator provides a stroboscopic description of chaotic
scattering in a quantum dot \cite{Osi03}, in the sense that the wave
function is determined only at times which are multiples of a time
$\tau_{0}$ (which we set to unity). The mean dwell time
$\tau_{D}=M/2N=\pi/N\delta$ is the ratio of the dimension $M$ of the
Floquet matrix (corresponding to a mean level spacing $\delta=2\pi/M$)
and the dimension $2N$ of the scattering matrix (without the dephasing
lead). The kicking strength $K=7.5$ determines the Lyapunov exponent
$\lambda=\ln(K/2)=1.32$. The Ehrenfest time is given by
\cite{Vav02,Sil03b}
\begin{equation}
\tau_{E}=\left\{\begin{array}{cc} \lambda^{-1}\ln(N^{2}/M)&{\rm
if}\;\;N>\sqrt{M},\\ 0&{\rm if}\;\;N<\sqrt{M}.
\end{array}\right.\label{tauEdef}
\end{equation}

The dephasing lead increases the dimension of the scattering matrix
$S$ to $M\times M$. It has the block form
\begin{equation}
S=\left(\begin{array}{ccc} s_{00}&s_{01}&s_{02}\\
s_{10}&s_{11}&s_{12}\\ s_{20}&s_{21}&s_{22}
\end{array}\right),\label{Sblock}
\end{equation}
where the subscripts $1,2$ label the two real leads and $0$ labels the
dephasing lead. The two real $N$-mode leads are coupled ballistically
to the system, while the remaining $M-2N$ modes are coupled via a
tunnel barrier. The dephasing rate $1/\tau_{\phi}=\Gamma(1-2N/M)$ is
proportional to the tunnel probability $\Gamma$ per mode. The
dephasing lead is connected to an electron reservoir at a voltage
which is adjusted so that no current is drawn. The conductance $G$ is
then determined by the coefficients $G_{ij}={\rm
Tr}\,s_{ij}^{\vphantom{\dagger}}s_{ij}^{\dagger}$ through
B\"{u}ttiker's formula \cite{But86},
\begin{eqnarray}
G&=&G_{12}+\frac{G_{10}G_{02}} {G_{10}+G_{20}}\nonumber\\
&=&G_{12}+\frac{(N-G_{11}-G_{12})(N-G_{22}-G_{12})}{2N-G_{11}-G_{12}-G_{22}-G_{21}}.
\label{G3leads}
\end{eqnarray}.

For $\Gamma\ll 1$ the dephasing lead model is equivalent to the
imaginary energy model of dephasing \cite{Bro97}, which is the model
used by Aleiner and Larkin \cite{Ale96}. (We will also make use of
this equivalent representation later on.) There exist other models of
dephasing in quantum transport \cite{Mar04,Cle04}, but for a
comparison with Ref.\ \cite{Ale96} our choice seems most appropriate.

Since we need a relatively small Lyapunov exponent in order to reach a
large enough Ehrenfest time, our simulations are sensitive to short
non-ergodic trajectories. These introduce an undesired dependence of
the data on the position of the leads. Preliminary investigations
indicated that UCF in a magnetic field was least sensitive to the lead
positions, so we concentrate on that quantum interference effect in
the numerics. The variance ${\rm Var}\,G$ of the conductance was
calculated in an ensemble created by sampling 40 values of the
quasi-energy. To determine the dependence on the lead positions we
repeated the calculation for 40 different configurations of the
leads. Error bars in the plots give the spread of the data.

\begin{figure}
\includegraphics[width=8cm]{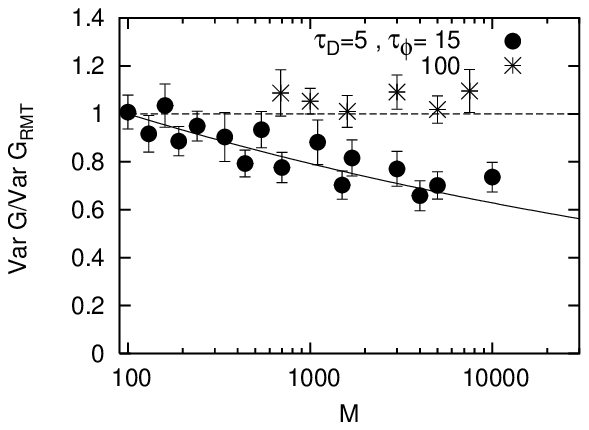}\\
\includegraphics[width=8cm]{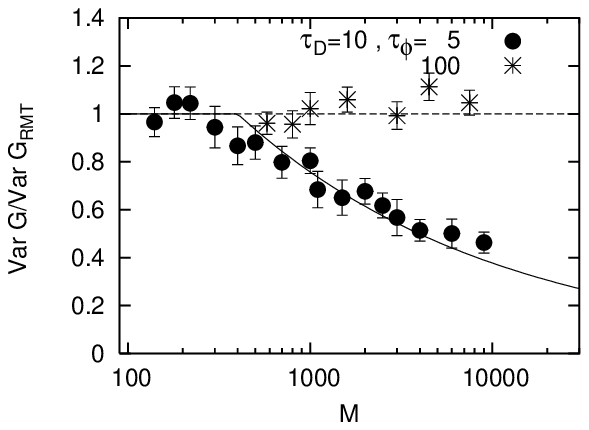}
\caption{ Variance of the conductance fluctuations, normalized by the
RMT value (\ref{VarGRMT}), as a function of the dimension $M$ of the
scattering matrix of the kicked rotator with a dephasing lead. Each
data set is for a fixed value of the dwell time $\tau_{D}=M/2N$ and
dephasing time $\tau_{\phi}=\Gamma^{-1}(1-2N/M)^{-1}$. The Lyapunov
exponent $\lambda=1.32$ is kept the same for all data sets. The curves
show the Ehrenfest time dependence (\ref{VarGresult}) predicted by the
effective RMT, without any fit parameter.
\label{varGplot}
}
\end{figure}

There are four time scales in the problem: $\lambda^{-1}$, $\tau_{D}$,
$\tau_{\phi}$, and $\tau_{E}$. To isolate the $\tau_{E}$ dependence we
increase both $M$ and $N$ at constant ratio $M/N$ and fixed
$K,\Gamma$. Then only $\tau_{E}$ varies. Results are shown in Fig.\
\ref{varGplot}. The variance of the conductance is divided by the RMT
prediction (\ref{VarGRMT}), with $\beta=2$ because of broken
time-reversal symmetry \cite{Note1}. We see that for
$\tau_{\phi}\gg\tau_{E}$ there is no systematic dependence of UCF on
the Ehrenfest time, consistent with Refs.\
\cite{Jac04,Two04a}. However, an unambiguous $\tau_{E}$-dependence
appears for $\tau_{\phi}\alt\tau_{E}$, regardless of whether
$\tau_{\phi}$ is smaller or larger than $\tau_{D}$.

\begin{figure}
\includegraphics[width=8cm]{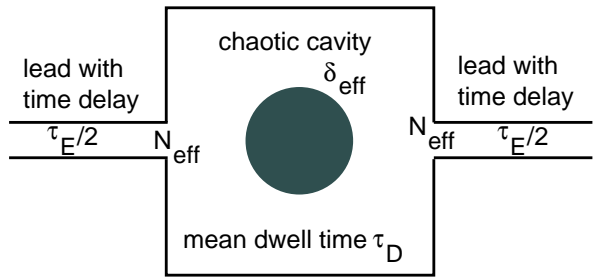}
\caption{Pictorial representation of the effective RMT of a ballistic
chaotic quantum dot. The part of phase space with dwell times
$>\tau_{E}$ is represented by a fictitious chaotic cavity (mean level
spacing $\delta_{\rm eff}$), connected to electron reservoirs by two
long leads ($N_{\rm eff}$ propagating modes, one-way delay time
$\tau_{E}/2$ for each mode). The effective parameters are determined
by $N_{\rm eff}/N=\delta/\delta_{\rm
eff}=\exp(-\tau_{E}/\tau_{D})$. The scattering matrix of lead plus
cavity is $\exp(iE\tau_{E}/\hbar)S_{\rm RMT}(E)$, with $S_{\rm
RMT}(E)$ distributed according to RMT. A finite dephasing time
$\tau_{\phi}$ is introduced by the substitution $E\rightarrow
E+i\hbar/2\tau_{\phi}$. The part of phase space with dwell times
$<\tau_{E}$ has a classical scattering matrix, which does not
contribute to quantum interference effects. \label{fictcavity} }
\end{figure}

To explain the data in Fig.\ \ref{varGplot} we introduce dephasing
into the effective RMT. For that purpose it is more convenient to use
an imaginary energy than a dephasing lead, so we first make the
connection between these two equivalent representations. There exists
an exact correspondence for any $N$ \cite{Bro97}, which requires a
re-injection step to ensure current conservation. For the case $N\gg
1$ of interest here there is a simpler way.

The coefficients $G_{ij}=G_{ij}^{\rm cl}+G_{ij}^{\rm q}$ in Eq.\
(\ref{G3leads}) consist of a classical contribution $G_{ij}^{\rm cl}$
of order $N$ plus a (sample specific) quantum correction $G_{ij}^{\rm
q}$ of order unity. The classical contribution is
\begin{equation}
G_{ij}^{\rm
cl}={\textstyle\frac{1}{2}}N(1+\tau_{D}/\tau_{\phi})^{-1},\;\;{\rm
for}\;\; i,j\in\{1,2\}. \label{Gijcl}
\end{equation}
Substitution into Eq.\ (\ref{G3leads}) gives a classical conductance
$G_{\rm cl}=N/2$ independent of dephasing --- as it should be. To
leading order in $N$ we obtain the quantum correction to the
conductance,
\begin{equation}
G_{\rm
q}={\textstyle\frac{1}{4}}(G_{12}+G_{21}-G_{11}-G_{22}). \label{deltaGrelation}
\end{equation}
(Notice that the classical contribution drops out of the
right-hand-side.) For $\Gamma\ll 1$ the effect of the dephasing lead
on the coefficients $G_{ij}$ is equivalent to the addition of an
imaginary part $i\hbar/2\tau_{\phi}$ to the energy. With the help of
Eq.\ (\ref{deltaGrelation}) we can compute the effect of dephasing on
WL and UCF,
\begin{eqnarray}
&&\Delta G=\langle G_{\rm
q}(E+i\hbar/2\tau_{\phi})\rangle,\label{DeltaGGq}\\ &&{\rm
Var}\,G=\langle \left[G_{\rm
q}(E+i\hbar/2\tau_{\phi})\right]^{2}\rangle-(\Delta G)^{2},
\label{VarGGq}
\end{eqnarray}
by averaging the scattering matrix at a complex energy without having
to enforce current conservation.

Effective RMT \cite{Sil03a} is a phenomenological decomposition of the
scattering matrix $S(t)$ in the time domain into a classical
deterministic part $S_{\rm cl}$ for $t<\tau_{E}$ and a quantum part
$S_{\rm q}$ with RMT statistics for $t>\tau_{E}$,
\begin{equation}
S(t)=\left\{\begin{array}{ll} S_{\rm cl}(t)&{\rm if}\;\;t<\tau_{E},\\
S_{\rm q}(t)=S_{\rm RMT}(t-\tau_{E})&{\rm if}\;\;t>\tau_{E}.
\end{array}\right. \label{Stdef}
\end{equation}
The RMT part $S_{\rm q}$ couples to a reduced number $N_{\rm
eff}=N\exp(-\tau_{E}/\tau_{D})$ of channels in each lead. The mean
dwell time in the quantum dot of these channels is
$\tau_{E}+\tau_{D}$. The classical part $S_{\rm cl}$ couples to the
remaining $2(N-N_{\rm eff})$ channels. (See Ref.\ \cite{Sil03b} for an
explicit construction of $S_{\rm cl}$.)

Only $S_{\rm q}$ contributes to $G_{\rm q}$. Fourier transformation to
the energy domain gives
\begin{equation}
S_{\rm q}(E)=e^{iE\tau_{E}/\hbar}S_{\rm RMT}(E),\label{Sqdef}
\end{equation}
where we have used that $S_{\rm RMT}(t)=0$ if $t<0$. The matrix
$S_{\rm RMT}$ has the RMT statistics of a fictitious chaotic cavity
with zero Ehrenfest time, $N_{\rm eff}$ modes in each lead, and the
same mean dwell time $\tau_{D}$ as the real cavity (see Fig.\
\ref{fictcavity}). For real energy the phase factor
$\exp(iE\tau_{E}/\hbar)$ is irrelevant, hence all
$\tau_{E}$-dependence is hidden in $N_{\rm eff}$ and $\delta_{\rm
eff}$. Since $\Delta G$ and ${\rm Var}\,G$ are independent of these
two parameters, they are also independent of $\tau_{E}$. The imaginary
part $i\hbar/2\tau_{\phi}$ of the energy that represents the dephasing
introduces a $\tau_{E}$-dependence of $G_{\rm
q}\propto\exp(-\tau_{E}/\tau_{\phi})$. Insertion of this factor into
Eqs.\ (\ref{DeltaGGq}) and (\ref{VarGGq}) yields the results
(\ref{DeltaGresult}) and (\ref{VarGresult}) given in the introduction.

\begin{figure}
\includegraphics[width=8cm]{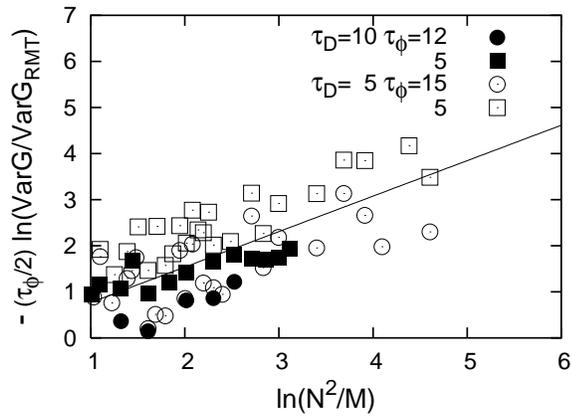}
\caption{ Four data sets of fixed $\tau_{\phi}$, $\tau_{D}$, each
consisting of a range of $M$ between $10^{2}$ and $2 \cdot 10^{4}$, plotted on
a double-logarithmic scale. The solid line with slope $1/\lambda=0.76$
is the scaling predicted by Eqs.\ (\ref{tauEdef}) and
(\ref{VarGresult}).
\label{loglogscaling}
}
\end{figure}

The curves in Fig.\ \ref{varGplot} follow from Eq.\
(\ref{VarGresult}). They describe the simulation quite well ---
without any fit parameter.  To test the agreement between simulation
and effective RMT in a different way, we have collected all our data
in Fig.\ \ref{loglogscaling} in a plot of $-(\tau_{\phi}/2)\ln({\rm
Var}\,G/{\rm Var}\,G_{\rm RMT})$ versus $\ln(N^{2}/M)$. According to
Eq.\ (\ref{VarGresult}) this should be a plot of $\tau_{E}$ versus
$\ln(N^2/M)$, which in view of Eq.\ (\ref{tauEdef}) is a straight line
with slope $1/\lambda=0.76$. There is considerable scatter of the data
in Fig.\ \ref{loglogscaling}, but the systematic parameter dependence
is consistently described by the theory as $N$ and $M$ vary over two
orders of magnitude.

In conclusion, our findings explain the puzzling difference in the
outcome of previous experimental \cite{Yev00} and numerical
\cite{Jac04,Two04a,Two04b} searches for the Ehrenfest time dependence
of quantum interference effects in chaotic systems: The experiments
found a dependence while the computer simulations found none. We have
identified the absence of dephasing in the simulations as the origin
of the difference. By introducing dephasing into the simulation we
recover the exponential $\tau_{E}/\tau_{\phi}$ suppression factor
predicted by Aleiner and Larkin \cite{Ale96}. The effective RMT
explains why this suppression factor is observed while the exponential
$\tau_{E}/\tau_{D}$ suppression factor of Eq.\ (\ref{deltaGAL}) is
not.

It remains an outstanding theoretical challenge to provide a
microscopic foundation for the effective RMT, or alternatively, to
derive Eqs.\ (\ref{DeltaGresult}) and (\ref{VarGresult}) from the
quasiclassical theory of Refs.\ \cite{Ale96,Ada03}. One might think
that diffraction of a wave packet at the point contacts is the key
ingredient which is presently missing from quasiclassics and which
would eliminate the exponential $\tau_{E}/\tau_{D}$ suppression factor
from Eq.\ (\ref{deltaGAL}). However, our observation of an exponential
$\tau_{E}/\tau_{\phi}$ suppression factor suggests otherwise: If
diffraction at the edge of the point contacts were the dominant
mechanism by which wave packets are split into partial waves, then the 
characteristic time scale for the suppression of quantum interference
by dephasing would not be different from the mean dwell time $\tau_{D}$.

This work is part of the research program of the Dutch Science
Foundation NWO/FOM. J.T. acknowledges support by the European
Community's Human Potential Program under contract
HPRN--CT--2000-00144, Nanoscale Dynamics. P.W.B. acknowledges support
by the Packard Foundation.

\end{document}